\documentclass[aps,prl,twocolumn,superscriptaddress,showpacs,floatfix]{revtex4-1}
\usepackage[english]{babel}
\usepackage{graphicx}
\usepackage{stmaryrd}
\usepackage{amssymb}
\usepackage{amsfonts}
\usepackage{amsmath}
\usepackage{color}
\usepackage{xspace}
\usepackage{multirow}

\usepackage{bm}

\usepackage{color}
\definecolor{darkblue}{rgb}{0.18, 0.185, 0.573}
\definecolor{lightblue}{rgb}{0.13, 0.26, 0.99}

\usepackage[dvipdfmx,
colorlinks=true,
urlcolor=lightblue,
citecolor=lightblue,
linkcolor=lightblue,
hyperfootnotes=false]{hyperref}

\DeclareGraphicsExtensions{.pdf,.ps,.eps}

\begin{document}

\title{
Boundary Resonances in $S=1/2$ Antiferromagnetic Chains under a Staggered Field}
\author{Shunsuke C. Furuya}
\affiliation{Department of Physics, University of Tokyo, Hongo, Tokyo
113-0033, Japan}
\affiliation{Institute for Solid State Physics, University of Tokyo,
Kashiwa 277-8581, Japan}
\author{Masaki Oshikawa}
\affiliation{Institute for Solid State Physics, University of Tokyo,
Kashiwa 277-8581, Japan}
\date{\today}

\begin{abstract}
We develop a boundary field theory approach to electron spin resonance
 in open $S=1/2$ Heisenberg antiferromagnetic  chains with an
effective staggered field.
In terms of the sine Gordon effective field theory with boundaries,
we point out the existence of boundary bound states of elementary
excitations, and modification of the selection rules at the boundary.
We argue that several ``unknown
 modes'' found in electron spin resonance experiments on
 KCuGaF$_6$~[I. Umegaki \textit{et al.}, Phys. Rev. B {\bf 79}, 184401
 (2009)] and Cu-PM~[S. A. Zvyagin \textit{et al.}, Phys. Rev. Lett. {\bf
 93}, 027201 (2004)] can be understood as boundary resonances
introduced by these effects.
\end{abstract}


\maketitle

\textit{Introduction. ---}
Impurities often introduce new aspects in physics,
Kondo effect being a notable example.
In particular, impurity effects in strongly correlated
systems are currently
among central topics in condensed matter physics.
Although the standard perturbation theory can fail,
there are number of powerful theoretical approaches to
strongly correlated systems, especially in one dimension.
Impurity effects in gapless one-dimensional systems
have been vigorously studied in terms of
boundary conformal field theory. 
In contrast, impurity effects in gapped one-dimensional systems
received much less attention,
with the exception of the edge states
in the $S=1$ Haldane gap
phase~\cite{hagiwara_endspin,batista_esr,yoshida_esr}.

On the other hand, integrable models and field theories
have been successfully applied to many gapped one-dimensional
systems.
In quantum magnetism, a field-induced gap in $S=1/2$
Heisenberg antiferromagnetic (HAFM) chains is described in
terms of a quantum sine Gordon field
theory~\cite{Dender,figap_prl,affleck_staghx};
the one-dimensional Ising chain with critical transverse
field and a weak longitudinal field realizes
a quantum field theory with $E_8$ symmetry~\cite{Zam_E8,Coldea_E8}.
Experimental studies indeed found
elementary excitations predicted by these integrable
field theories.
Application of integrable field theories to
impurity/boundary effects in gapped one-dimensional
strongly correlated systems is an interesting
but largely unexplored subject.

In this Letter, we present a theory of
electron spin resonance (ESR) in $S=1/2$ HAFM chains in a staggered field with
boundaries, which may be realized by nonmagnetic impurities.
The low-energy effective theory of the system
is the quantum sine Gordon field theory with boundaries.
In fact, this theory is integrable even in the presence of a boundary,
and boundary bound states (BBS) of elementary excitations
have been found in the exact
solution~\cite{ghoshal_smatrix,ghoshal_bsg,mattsson_boundaryspec,bajnok_neumann_bsg,Schuricht_PRL,schuricht_boundaryLDOS}.
The existence of BBS, and modification of the selection rules,
imply extra resonances in addition to those in the bulk.
ESR measurements on corresponding systems
KCuGaF$_6$~\cite{umegaki_KCuGaF6} and 
[PM-Cu(NO$_3$)$_2$(H$_2$O)$_2$]$_n$ (PM denotes pyrimidine, abbreviated as
Cu-PM)~\cite{zvyagin_CuPM_prl,zvyagin_CuPM}
had found several resonances that could not be accounted for
by the theory.
We argue that several of those resonances can be successfully
identified in terms of the boundary sine Gordon field theory.

\textit{Boundary sine-Gordon field theory. ---}
We consider a semiopen chain with the Hamiltonian
\begin{equation}
  \mathcal H = \sum_{j < 0} \bigl[ J \bm S_j \cdot \bm S_{j-1}
   -\mu_B\bm H \cdot {\bf g} \cdot \bm  S_j 
  +(-1)^j \bm D \cdot \bm S_j \times \bm S_{j-1}\bigr],
  \label{eq.Hini}
\end{equation}
which models one side of an infinite chain
broken by a nonmagnetic impurity at $j=0$.
${\bf g}$ is the $g$ tensor of
localized spins, and $\mu_B$ is Bohr magneton.

The Zeeman energy can be represented as
$-\mu_B\sum_{j,a,b}H^a\bigl[g^u_{ab} + (-1)^j g^s_{ab}\bigr]S^b_j$.
Hereafter, we assume that $g^u_{ab}  g \delta_{ab}$ and
$|g^s_{ab}| \ll g$, and
employ a unit $\hbar = k_B = g\mu_B = 1$.

We consider $\bm H = H \hat z$ applied along the $z$ direction ($\hat z$ is
a unit vector in the $z$ direction).
The last term of \eqref{eq.Hini} is 
the staggered Dzyaloshinskii-Moriya (DM) interaction.
This can be eliminated by a staggered rotation of spin 
about the direction of $\bm D$ by angle $(-1)^j \alpha/2$,
where $\alpha = \tan^{-1}(|\bm D|/J)$ where $j$ is the site index.

Under an applied field, this transformation leaves
a staggered field $\bm h \sim \bm H \times \bm D /2$,
which is perpendicular to $\bm H$
and to $\bm D$.~\cite{figap_prl,affleck_staghx}
Together with the staggered field due to the staggered component
of the $g$ tensor, the effective model may be given by
\begin{equation}
 \mathcal H = \sum_{j < 0} \bigl[ J \bm S_j \cdot \bm S_{j-1}
  -HS^z_j - h(-1)^j S^x_j \bigr],
  \label{eq.H}
\end{equation}
keeping only the most important terms.
$h = c_s H$ is the effective staggered field, approximately perpendicular
to the applied field; the direction of the staggered field is chosen to be
the $x$ axis.
$c_s$ depends both on the staggered DM interaction
and on the staggered part of $g$ tensor $g^s_{ab}$.
Using bosonization formulas,
\begin{align}
 S^z_x &\sim m + \frac 1{2\pi R} \partial_x \phi + C^z_s(-1)^x \cos
 (\phi/R + Hx),
 \label{eq.Sz2phi} \\
 S^\pm_x &\sim e^{\pm  2\pi R i  \tilde \phi} \bigl[ C^\perp_s (-1)^x
 + C^\perp_u \cos (\phi/R + Hx)\bigr],
 \label{eq.Spm2phi}
\end{align}
at low temperature $T \ll J$, the model \eqref{eq.H} is mapped to a
boundary sine Gordon (BSG) field theory, defined by the action
\begin{multline}
 \mathcal A = \int_{-\infty}^\infty dt\int_{-\infty}^0dx \biggl[ \frac
  12 \biggl\{\frac 1{v^2}(\partial_t \tilde \phi)^2 - (\partial_x \tilde
  \phi)^2\biggr\} \\
 - C^\perp_s h \cos{(2\pi R \tilde \phi)}\biggr].
  \label{eq.A}
\end{multline}
$v$ is the spin-wave velocity.
The fields $\phi$ and $\tilde \phi$ are 
dual and
compactified as $\phi \sim \phi
+ 2\pi R$ and $\tilde \phi \sim \tilde \phi +1/R$;
$m$ is the uniform magnetization density and
the nonuniversal constants $C^z_{u,s}$ and $C^\perp_{u,s}$
are determined numerically~\cite{hikihara_coefficients}.
The bulk operator $\cos{(2 \pi R \tilde{\phi})}$, which
represents the transverse staggered magnetization, is relevant in the 
renormalization group sense. 
Thus it induces a finite mass (excitation gap), as it
was observed in the experiments.~\cite{figap_prl,affleck_staghx}
The bulk gap is stable against dilute nonmagnetic impurities.

In the absence of the staggered field term,
the field $\phi$ obeys the Dirichlet boundary condition
$\phi(x=0,t)=\mathrm{const.}$~\cite{eggert_imp}.
This is equivalent to the Neumann boundary condition
in terms of the dual field:
\begin{equation}
 \left.\partial_x \tilde \phi(x,t)\right|_{x=0}=0.
  \label{eq.Neumann}
\end{equation}
Inclusion of the staggered field could change the boundary
condition; in fact, the staggered field in the bulk
would also induce the corresponding operator
$\cos{(2\pi R\tilde{\phi})}$ at the boundary.
If this boundary perturbation is dominant, 
$\tilde{\phi}$ would obey the Dirichlet boundary condition.
However, since $\cos{(2 \pi R \tilde{\phi})}$ is,
as a boundary operator,
(nearly) marginal in RG, its effects are negligible
at the energy scale set by the bulk spin gap.
Thus, for a small staggered field $h$,
the boundary condition can still be regarded as the
Neumann on $\tilde{\phi}$.

\textit{Energy spectrum. ---}
The elementary excitation of the bulk sine Gordon field theory
includes soliton (denoted by $S$) and antisoliton
($\bar S$) with the same mass $M$.
A soliton and antisoliton carry soliton charge
$Q=+1$ and $Q=-1$, respectively.
Additional particles called breathers are
generated as bound states of
a soliton and an antisoliton~\cite{DHN,Zam2}.
There can be several different kinds of breathers $B_n$
with the mass $ M_n = 2M \sin (n\pi \xi/2)$,
for $n=1,2,\ldots, \lfloor \xi^{-1}\rfloor$.
Here $\lfloor \cdot \rfloor$ is the floor function.
Breathers have zero soliton charge.
The soliton charge is conserved in the bulk sine Gordon field theory.
For the case of our interest, that is experimental situations in
KCuGaF$_6$ and Cu-PM, the parameter $\xi = 1/(2/\pi R^2 - 1)$
satisfies $\xi < 1/3$.
In particular, $\xi \approx 1/3$ in the low field limit $H \to 0$.

Fortunately, the BSG theory \eqref{eq.A} with
the Neumann boundary condition \eqref{eq.Neumann}
is still integrable~\cite{ghoshal_smatrix}.
An analysis of the \emph{boundary} $S$ matrices
implies~\cite{ghoshal_smatrix,ghoshal_bsg,mattsson_boundaryspec,bajnok_neumann_bsg,schuricht_boundaryIsing}
the existence of the BBS with the mass
\begin{equation}
 M_{\mathrm{BBS}} = M \sin (\pi \xi),
  \label{eq.M_BB}
\end{equation}
for $\xi<1/2$.
Therefore, the BBS with $M_{\mathrm{BBS}} \sim \sqrt{3}M/2$ does exist
in the low field limit of the present system.
The BBS of soliton, antisoliton, and first breather
turn out to be identical and there is only one type of BBS.
This is another manifestation of the soliton charge
nonconservation at the boundary.

\begin{figure}[b!]
 \centering
 \includegraphics[width=\linewidth]{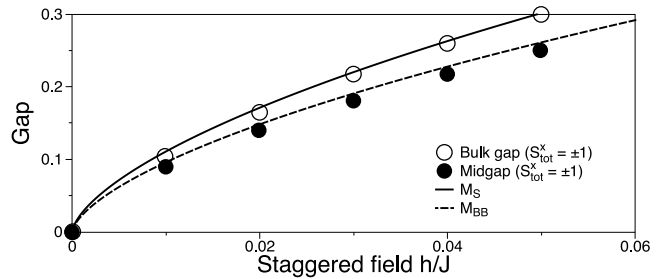}
 \caption{Numerically obtained bulk gap (open circles) and midgap (solid
 circles) in \cite{lou_midgap_prl} are compared with
 the soliton mass $M$ and the BBS mass $M_{\mathrm{BBS}}$ \eqref{eq.M_BB}.}
 \label{fig.midgap}
\end{figure}

Thus the analysis of the BSG theory predicts a new excited state,
which is a BBS, at the energy $M_{\mathrm{BBS}}$
lower than the bulk gap $M$.
In fact, in an earlier numerical study of an open chain based on 
density matrix renormalization group,
Lou \textit{et al.}~\cite{lou_midgap_prl,lou_midgap_prb}
had found such an excited state localized
near the boundary. They called it a midgap state.
Figure~\ref{fig.midgap} shows a comparison of the soliton~\eqref{eq.M}
and BBS~\eqref{eq.M_BB} masses,
to the numerically obtained bulk gap and the energy of
the midgap state in Refs.~\cite{lou_midgap_prl,lou_midgap_prb}.
Here we used
$M\approx 1.85\, (h/J)^{2/3}[\ln (J/h)]^{1/6}$
appropriate for $H = 0$, which was used in
Refs.~\cite{lou_midgap_prl,lou_midgap_prb}.
The excellent agreement between the two energies means that
the midgap state found in the
DMRG calculation was nothing but a BBS.

While it was pointed out in
Refs.~\cite{lou_midgap_prl,lou_midgap_prb}
that the midgap state is localized near the boundary,
physical understanding of its origin has been lacking.
The analogy to the edge state in the Haldane phase
is clearly inappropriate, since the $S=1/2$ HAFM in
a staggered field is topologically trivial.
With the present identification of the midgap state with
the BBS, the BSG theory is established as an effective
theory describing the boundary physics of the system.
This is essential in understanding the ESR spectra.

\textit{Electron spin resonance. ---}
Next we discuss the ESR spectrum, which is given by
$I(\omega) \propto \omega \chi''_{\mathrm{phys}}(q=0,\omega)$.
Here, $\chi''$ is an imaginary part of the dynamical
susceptibility $\chi$.
$\chi_{+-}(q, \omega) = -G_{S^+S^-}(q, i\omega = \omega + i\epsilon)$
where positive infinitesimal $\epsilon$ is the
analytic continuation of the temperature Green's function
$G_{S^+S^-}(q, i\omega)$.
The staggered rotation of spins by angles $(-1)^j \alpha/2$ to
eliminate the DM interaction mix the uniform ($q=0$)
and the staggered ($q=\pi$) components.
The physical susceptibility $\chi''_{\mathrm{phys}}(q=0,\omega)$ is
\begin{align}
 \chi''_{\mathrm{phys}}(0, \omega) 
 &\sim \chi''_{+-}(0, \omega)
 + \biggl(\frac{D_z} J \biggr)^2
 \chi''_{+-}(\pi, \omega) 
 \label{eq.chi_phys},
\end{align}
where $D_z$ is the $z$ component of the DM vector $\bm D$
parallel to the applied field.
On the right-hand side, we dropped
the longitudinal susceptibility $\chi_{zz}(q=\pi, \omega)$
because it contains the same resonances as those
in the transverse part $\chi_{+-}(q=0, \omega)$, and
merely modifies their intensities.

Let us first review the ESR in the bulk, in the limit
of $T \rightarrow 0$.
The uniform part $\chi''_{+-}(0,\omega)$ reflects
transitions caused by the operator
\begin{equation}
 S^\pm_{q=0} \sim e^{\pm 2\pi R i \tilde \phi} \cos{(\phi/R + Hx)} . 
\label{eq.Spm.unif}
\end{equation}
The operator $\cos{(\phi/R)}$ changes the soliton charge
by $\pm 1$, and thus must create at least one
soliton or antisoliton.
The other factor $e^{\pm 2\pi R i \tilde \phi}$ can create
any number of excitations with zero soliton charge in total.
Thus ESR induced by the operator~\eqref{eq.Spm.unif} with
the lowest possible energy corresponds to creation
of a single soliton or antisoliton.
It should be also noted that
the factor $Hx$ in the cosine causes
the shift of the momentum with $H$.
That is, the created soliton or antisoliton
should carry the momentum $H$.
Thus the ESR due to a single soliton/antisoliton
creation is at frequency
$\omega = E_S \equiv \sqrt{M^2 +H^2}$.
There are also resonances due to~\eqref{eq.Spm.unif}
at higher energies,
corresponding to creation of additional elementary
excitations. 

Next we turn to the staggered part $\chi''_{+-}(q=\pi,\omega)$,
which reflects transitions caused by the operator
\begin{equation}
 S^\pm_{q=\pi} \sim e^{\pm 2\pi R i \tilde \phi} .
\label{eq.Spm.stag}
\end{equation}
This carries zero soliton charge, and thus the
simplest excitation induced by this operator
is creation of a single breather $B_n$.
This leads to the resonances at $\omega = M_n$.

Now let us discuss the boundary effects on ESR.
Here we discuss the ESR spectrum based on physical picture,
leaving systematic formulations to the Supplementary Material~.

First we consider the contribution of the
staggered part $\chi''_{+-}(q=\pi, \omega)$.
The simplest excitation created by the
operator~\eqref{eq.Spm.stag} is a single breather.
In the presence of the boundary, the first breather
can form the BBS.
Thus the resonance with the lowest energy in the
presence of the boundary is given by
$\omega = M_{\mathrm{BBS}}$.
Creation of a breather not bounded at the boundary and the BBS is also
possible, leading to the resonance at $\omega = M_{\mathrm BBS} + M_n$.

In order to understand the boundary effects on
ESR, we need to clarify the issue of
the momentum conservation.
In general, total momentum is conserved
due to the translation invariance of the system.
In the presence of the boundary, the translation invariance
is lost and the total momentum is no longer conserved.
Nevertheless, the momentum is still important in
discussing ESR spectra, because the momentum of each
elementary excitation is conserved or reversed
in a scattering with another elementary excitation,
or in a reflection at the boundary.
Thus, once created, the set of momenta of elementary
excitations is conserved, up to the sign of each momentum.

\begin{table}[t!]
\begin{tabular}[t]{ccc|c|c}
 \hline
 \hline
 &&  &Bulk&  Boundary \\
 \hline
 &$T=0$&
 & $\omega = E_S, E_S + M_n$ & $\omega = E_n, E_S+M_{\mathrm{BBS}}, E_n + M_{\mathrm{BBS}}$\\
 \hline
 &$T>0$& & $\omega = |E_S-M_n|$ &
 $\omega =
E_n -M_{\mathrm{BBS}}$ \\
 \hline
 \hline
\end{tabular}
 \caption{Typical resonance modes in $\chi''_{+-}(q=0, \omega)$.
 Resonances shown in the second row are absent at $T=0$.
 The soliton resonance $E_S$ is accompanied with all bulk resonances.
 On the other hand, $E_S$ does not necessarily appear in the boundary
 resonances.
 In fact, some boundary resonances are involved with a novel resonance
 $E_n$ instead of $E_S$.
 }
 \label{table.0}
\end{table}
\begin{table}[t!]
\begin{tabular}{ccc|c|c}
 \hline
 \hline
 &&& Bulk & Boundary \\
 \hline
 &$T=0$&
 & $\omega = M_n, M_n + M_m$ & 
	 $\omega = M_{\mathrm{BBS}}, M_n + M_{\mathrm{BBS}}$\\
 \hline
 &$T>0$& & $\omega = M_n - M_m$ ($n>m$) & $\omega = M_n -
	 M_{\mathrm{BBS}}$ \\
 \hline \hline
\end{tabular}
 \caption{Typical resonance modes in $\chi''_{+-}(q=\pi, \omega)$.
 Breather masses are directly measured in the bulk
 part.
 The BBS mass itself appears in the boundary resonances.
 Equations~\eqref{eq.chi_phys} shows that intensities of these modes are smaller
 than those in Table~\ref{table.0} by $(D_z/J)^2$.
 }
 \label{table.pi}
\end{table}

\begin{figure*}
 \centering
 \includegraphics[width=\linewidth]{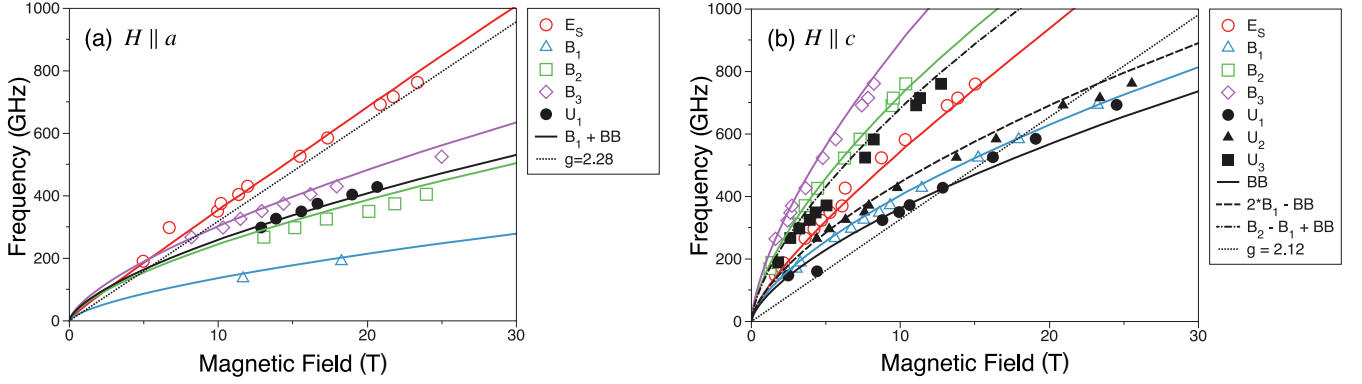}
 \caption{(color online) Frequency vs field diagrams of ESR in
 KCuGaF$_6$~\cite{umegaki_KCuGaF6}
 for (a) $H \parallel a$ and (b) $H \parallel c$ configurations.
 The dotted line is high temperature paramagnetic resonance $\omega
 = H$.
 Open symbols and filled symbols represent bulk modes and
 ``unknown modes.''
 $\mathrm{E_S, B_1, B_2, B_3}$ denotes resonances $\omega = E_S,
 M_1, M_2, M_3$ by a soliton $S$ and breathers $B_n$.
 An antisoliton $\bar S$ also leads to $\omega = E_S$.
 The labels $U_1, U_2, U_3$ are ``unknown'' peaks found
 in~\cite{umegaki_KCuGaF6}.
 (a) The configuration $H \parallel a$  bears the smallest $h=c_sH$
 with the coefficient $c_s = 0.031$.
 An excitation $\omega = M_1 + M_{\mathrm{BBS}}$ is found in addition
 to the bulk excitations.
 (b) We have the largest staggered field, $c_s = 0.178$, when $H
 \parallel c$.
 This large $c_s$ makes the rich kinds of boundary modes detectable.
 The labels BB, 2*B$_1$-BB, B$_2$-B$_1$+BB denote
 $\omega = M_{\mathrm{BBS}}, 2M_{1} - M_{\mathrm{BBS}}, M_{2} - M_{1} +
 M_{\mathrm{BBS}}$.
 }
 \label{fig.umegaki}
\end{figure*}

The existence of the boundary
has an interesting effect, in addition to the contribution
of the BBS, on the ESR spectrum.
Although the expectation of the operator $\cos{(\phi/R)}$
vanishes in the bulk, 
it is nonvanishing~\cite{FujimotoEggert,FujimotoEggert-Erratum}
near a boundary with the Dirichlet boundary condition on $\phi$
[equivalent to the Neumann boundary condition~\eqref{eq.Neumann}
on $\tilde{\phi}$]. 
This reflects $\phi$ taking a fixed value at the boundary.
Thus, \emph{in the vicinity of the boundary},
the leading contribution from the uniform part
is effectively given by the operator
$S^\pm_{q=0} \sim e^{\pm 2\pi R i \tilde \phi} \cos{(Hx)}$,
which creates excitations with zero soliton charge, in
contrast to the original one~\eqref{eq.Spm.unif}.
This is another consequence of violation of
soliton charge conservation at the boundary.
The created excitations should carry the total
momentum $\pm H$, up to the uncertainty $\sim 1/l_p$.
Here $l_p$ is the pinning lengths cale, namely
$\phi(x) \sim \mbox{const.}$ if $|x| < l_p$.

Thus, the simplest among the possible ESR processes
due to this operator is the creation of a single breather $B_n$
with momentum $\pm H$. This corresponds to the frequency
\begin{equation}
 E_n = \sqrt{{M_n}^2+H^2} .
\label{eq.EBn}
\end{equation}
The resonance at $E_n$ with $n>1$
was absent in the bulk and is a new feature due to
the boundary.
We emphasize that, these new resonances
do \emph{not} simply follow from the
existence of the midgap state numerically
found in Refs.~\cite{lou_midgap_prl,lou_midgap_prb}.
In fact, the resonance frequency $E_n$ does not
explicitly contain the energy $M_{\mathrm{BBS}}$.
This shows the necessity of
the BSG framework to fully understand the
physics at the boundary.

At finite temperatures,
additional resonances may be observable.
When the initial state contains the BBS as a thermal
excitation, a resonance at
$
 \omega = M_n - M_{\mathrm{BBS}}
$
exists, corresponding to
annihilation of the BBS and creation of a breather $B_n$.
Similarly, when the initial state contains $B_1$,
the resonance at
$
 \omega = M_n - M_1 + M_{\mathrm{BBS}}
$
corresponds to the creation of $B_n$ and binding of $B_1$
at the boundary.
These resonances are contained in the
staggered part $\chi''_{+-}(\pi,\omega)$.
The uniform part also contains, at finite temperatures,
additional resonances. 

Typical resonance modes are summarized in Tables~\ref{table.0} and
\ref{table.pi}.
Note that intensities of resonances due to the
staggered part $\chi''_{+-}(q=\pi,
\omega)$ and $\chi''_{zz}(q=\pi, \omega)$
are suppressed by the factor $(D_z/J)^2$ in \eqref{eq.chi_phys}, 
compared to those from the uniform part $\chi''_{+-}(q=0, \pi)$.

\begin{figure}[b!]
 \centering
 \includegraphics[width=\linewidth]{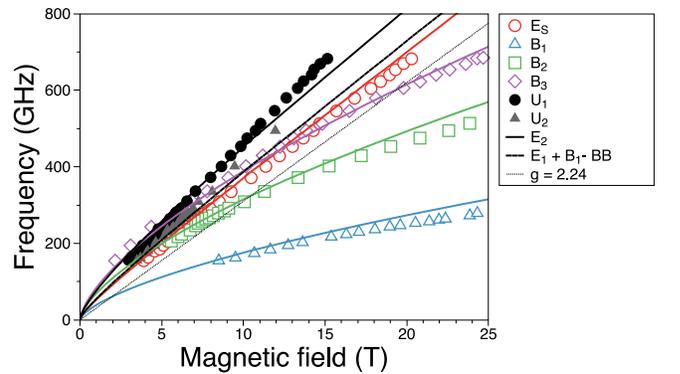}
 \caption{(color online) Frequency vs field diagrams of ESR in
 Cu-PM~\cite{zvyagin_CuPM_prl}.
 Two ``unknown'' peaks $U_1$ and $U_2$
 are attributed to
 $\omega = E_{2}, \, E_{1} +M_1 - M_{\mathrm{BBS}}$
 respectively, where $E_n$ is defined in Eq.\protect\eqref{eq.EBn}.
 These boundary modes appear in $\chi''_{xx}(q=0, \omega)$.
 }
 \label{fig.zvyagin}
\end{figure}

\textit{Comparison with experiments. ---}
Thus several novel resonances, which are absent in the bulk,
are derived from the BSG theory.
They indeed match the ``unknown'' resonances observed
previously in ESR spectra on KCuGaF$_6$ (Figure~\ref{fig.umegaki})
and Cu-PM (Figure~\ref{fig.zvyagin}).
Here the uniform field effect is taken into account in
the soliton-antisoliton mass~\cite{Lukyanov1997571},
\begin{equation}
 M = \frac{2v}{\sqrt \pi}\frac{\Gamma(\xi/2)}{\Gamma((1+\xi)/2)}
  \biggl[\frac{\Gamma(1/(1+\xi))}{\Gamma(\xi/(1+\xi))} \frac{C^\perp_s
  \pi }{2v} c_sH\biggr]^{(1+\xi)/2}.
  \label{eq.M}
\end{equation}
While the ratio $c_s = h/H$ can in principle be
determined by the staggered DM interaction and
$g$ tensor,
we use the value obtained by fitting experimental data.

Each figure shows different resonances, though.
The difference between the spectra in two materials
can be understood in terms of the different magnitude
of the DM interaction.
The coefficient $c_s$ at its maximum is larger
($0.178$) in KCuGaF$_6$  compared to $0.083$ in Cu-PM.
While precise estimates  of the DM interaction and 
the staggered $g$ tensor are not available,
the staggered DM interaction is presumably larger in KCuGaF$_6$.
This leads to larger mixing of the
staggered part $\chi''_{+-}(q=\pi,\omega)$.
Thus it is natural that
the resonances due to the mixings are observed
only in KCuGaF$_6$ (Fig.~\ref{fig.umegaki}). 
We note that, the simplest possible BBS contribution
at $\omega=M_{\mathrm{BBS}}$ is not observed
for $H \parallel a$ in Fig.~\ref{fig.umegaki}~(a).
This is presumably because only two of the frequencies
used in the experiments can detect the resonance
below the bulk resonance at $\omega=M_1$.
On the other hand, in the Cu-PM with a smaller
DM interaction, only the contributions from
the uniform part are observed (Fig.~\ref{fig.zvyagin}).
We conjecture that, with more careful examination of
the spectra, more resonances due to the BBS will be
found in experiments.

\textit{Conclusions. ---}
We point out the existence of BBS and modification of the selection
rules at boundaries of the $S=1/2$ antiferromagnetic chain
in an effective staggered field,
in terms of the BSG theory \eqref{eq.A}.
The boundary effects can account for the mysterious ``unknown modes''
found in two compounds KCuGaF$_6$~\cite{umegaki_KCuGaF6} and
Cu-PM~\cite{zvyagin_CuPM_prl,zvyagin_CuPM}.

In the compound KCuGaF$_6$,
magnetic ions Cu$^{2+}$ and nonmagnetic ions Ga$^{3+}$
form a pyrochlore lattice.
Magnetically, the compound KCuGaF$_6$ is effectively
regarded as $S=1/2$ HAFM chains of Cu$^{2+}$ ions.
However, since Cu$^{2+}$ and Ga$^{3+}$ ions occupy equivalent positions,
an intersite mixing of them can occur in the course of syntheses~\cite{mixing}.
We speculate that the intersite mixing brings about nonmagnetic
impurities in spin chains.
We expect that a few percent of the nonmagnetic impurities 
would lead to observation of the boundary resonances
as discussed in this Letter.
Our picture may be verified experimentally by controlling the
density of nonmagnetic impurity, which will change the intensity of
boundary resonances.

The authors thank P. Bouillot, T. Giamarchi, S. Takayoshi, H. Tanaka,
I. Umegaki, and S. Zvyagin for fruitful discussions.
This work is partly supported by the Global COE Program ``The Physical
Sciences Frontier'' (S.C.F.), and KAKENHI Grant No. 50262043
from MEXT of Japan.

%





\begin{table*}[t!]
{\bf \large Supplemental Material to \\
``Boundary Resonances in $S=1/2$ Antiferromagnetic Chains under
 a Staggered Field''}
\end{table*}

\newpage

\section{Effective lattice model}

\begin{figure}[h!]
 \centering
 \includegraphics[width=0.9\linewidth]{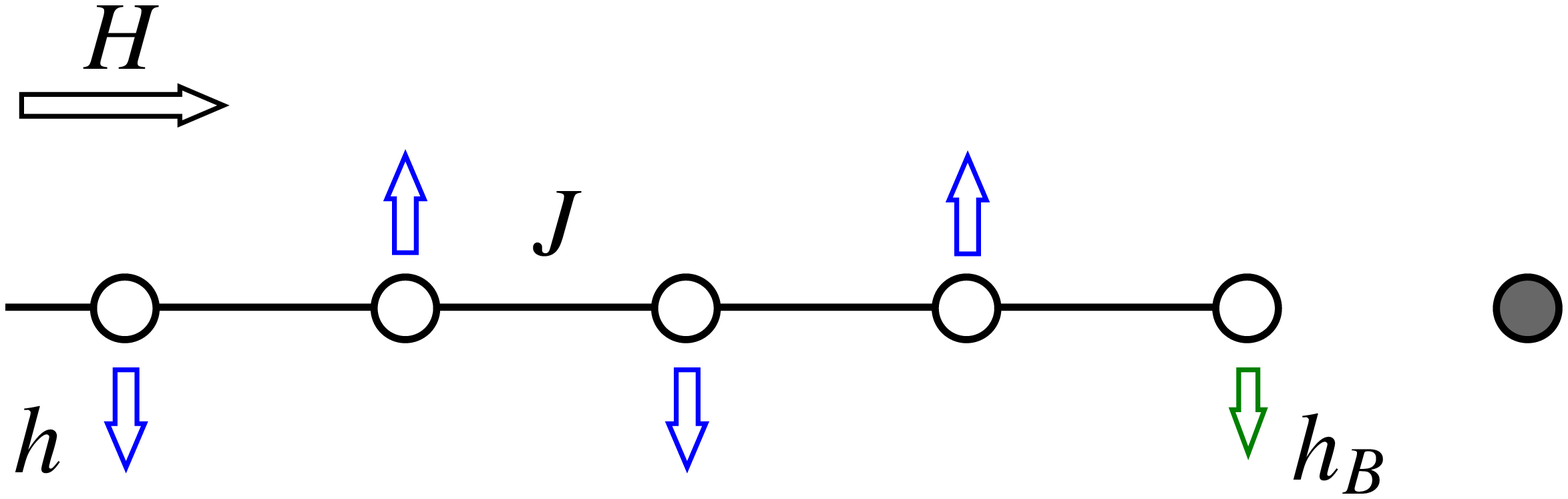}
 \caption{Semi-infinite Heisenberg antiferromagnetic chain with an
 effective staggered field, 
 cut by a non-magnetic impurity (the gray
 filled circle).}
\label{fig.chain}
\end{figure}

We derive the effective lattice model (2) of the main text, with
an emphasis on the mixing of $q=\pi$ modes
through the transformation eliminating the DM interaction.
First we consider the model Hamiltonian
\begin{equation}
 \mathcal H_0 = \sum_{j \le 0} \bigl[J \bm S_j \cdot \bm S_{j-1} - HS^z_j
  +(-1)^j \bm D \cdot \bm S_j \times \bm S_{j-1}\bigr].
  \label{eq.H0}
\end{equation}
Here, for simplicity, the effects of the anisotropic $g$ tensor
are ignored, and they will be taken into account later.
We rotate the spin $\bm S_j$ about the direction of $\bm D$
by the angle $(-1)^j\alpha/2$ where
$\alpha = \tan^{-1}(|\bm D |/J)$.
When $|\bm D| \ll J$, $\alpha \ll 1$ and the transformation
of the spin operator can be linearized as
\begin{equation}
  {\bm S}_j \to {\bm S}_j + (-1)^j \frac{\bm D}{2J} \times \bm S_j .
\label{eq.transform}
\end{equation}
Thus, in the presence of the uniform applied field $\bm H$,
a staggered field perpendicular to both $\bm H$ and $\bm D$ is
generated.

In addition, the staggered component of $g$ tensor 
also contributes to generating the staggered field.
In contrast to the one due to the staggered DM interaction, 
the staggered field coming from the staggered component of the
$g$ tensor is not necessarily perpendicular to
the applied field.
However, in the effective field theory description,
the longitudinal component of the staggered field
accompanies an oscillating factor, which makes it vanish
upon spatial integration.
Thus it can be ignored for the purpose of discussion of
gap opening, and only the transverse staggered field
has to be considered.

Therefore, the $S=1/2$ antiferromagnetic chain under an applied
magnetic field, with both staggered $g$ tensor and staggered DM
interaction  can be described by the effective lattice model (2)
of the main text. The effective staggered field $h$ depends on both
the staggered DM interaction and the staggered $g$ tensor.

In the course of eliminating the staggered DM interaction,
the spin operator is transformed according to
Eq.~\eqref{eq.transform}.
This has to be taken into account in the discussion
of physical susceptibility.
Namely, the physical uniform susceptibility $\chi''_{\mathrm{phys}}(q=0,
\omega)$ has contributions from both uniform and staggered
susceptibilities calculated for the effective model after the
elimination of the DM interaction: 
\begin{multline}
 \chi''_{\mathrm{phys}}(q=0, \omega) \sim \chi''_{+-}(q=0, \omega)\\
  + \biggl( \frac {D_z}J\biggr)^2 \chi''_{+-}(q=\pi, \omega)
 +\biggl( \frac{D_\perp}J\biggr)^2 \chi''_{zz}(q=\pi, \omega),
 \label{eq.mixing}
\end{multline}
where $D_\perp$ is the magnitude of the component $\bm D$
perpendicular to the $z$ axis.
As we will see below, the longitudinal susceptibility
$\chi''_{zz}(q=\pi, \omega)$
adds no resonant peak to the former two susceptibilities.
The longitudinal staggered susceptibility just modifies intensities
of the transverse susceptibilities.

Fig.~\ref{fig.chain} shows a schematic picture of the staggered fields
generated from the external uniform field $H$, in the bulk ($h$) and at
the boundary ($h_B$).
Note that the boundary staggered field $h_B$, coupled to 
the end spin $S^x_0$, is indistinguishable to the bulk
staggered field $h$ in the lattice model.
But their behaviors under renormalization group (RG) transformations are
quite different.

\section{RG analysis}

\begin{figure}
 \centering
 \includegraphics[width=\linewidth]{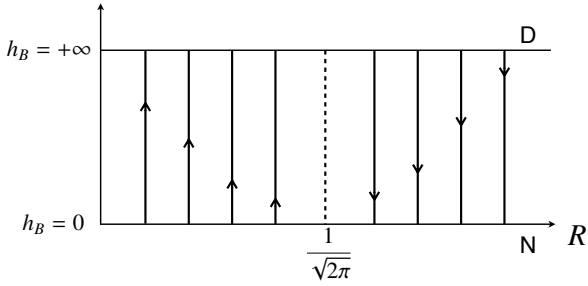}
 \caption{The boundary RG flow.
 On the line $h_B = 0$ denoted by ``N'', the boundary condition is
 Neumann.
 On the line $h_B = + \infty$ denoted by ``D'', the boundary condition
 is Dirichlet.
 }
 \label{fig.RG}
\end{figure}

The bosonization of spins leads to an effective boundary field theory
with an action,
\begin{align}
 \mathcal A 
 &= \int_{-\infty}^\infty dt \int_{-\infty}^0 dx\,
 \biggl[ \frac 12 \biggl\{
 \frac 1{v^2}(\partial_t \tilde \phi)^2 - (\partial_x \tilde
 \phi)^2\biggr\} \notag \\
 & \quad - C_s^\perp h \cos (2\pi R \tilde \phi)\biggr]
 -C_s^\perp h_B \int_{-\infty}^\infty dt \, \cos \bigl[ 2\pi R \tilde
 \phi (x=0)\bigr].
\end{align}
Scaling dimensions of the bulk and boundary staggered magnetizations are
respectively
\begin{equation}
 x_h = \pi R^2, \qquad x_{h_B} = 2\pi R^2.
  \label{eq.dimension}
\end{equation}
When the uniform field $H$ is zero, the effective lattice model (2) in
the paper is SU(2) symmetric, where the compactification radius $R$
is given by $R = 1/\sqrt{2\pi}$.
As the uniform and staggered fields increase, the compactification
radius decreases~\cite{affleck_staghx_suppl}.
Recall that a bulk operator is relevant (irrelevant) when its scaling
dimension is smaller (bigger) than 2,
and a boundary operator is relevant (irrelevant) when its scaling
dimension is smaller (bigger) than 1.
Thus, both the bulk ($x_h \le 1/2$) and boundary ($x_{h_B} \le 1$) 
staggered magnetizations are relevant.
The boundary staggered magnetization is marginal at zero field $H=0$.
The boundary RG flow is shown in Fig.~\ref{fig.RG}.
The line $h_B = 0$ corresponds to the Neumann boundary condition.
In the presence of the boundary staggered field, the boundary condition
is neither Neumann nor Dirichlet:
\begin{equation}
 \left. \partial_x \tilde \phi(x,t) \right|_{x=0}
  = 2\pi R C_s^\perp h_B \sin\bigl[2\pi R \tilde \phi(x=0, t)\bigr]
  \label{eq.bc}
\end{equation}
$h_B = + \infty$ corresponds to the Dirichlet boundary condition
because the field $\tilde \phi(x=0, t)$ is pinned to a certain value in
order to optimize the potential energy $\propto 
h_B \cos[2\pi R \tilde \phi(x=0,t)]$.

As we perform the RG transformation iteratively,
the bulk staggered field $h$ grows rapidly.
When the effective coupling
$h/J$ becomes $\mathcal O(1)$, 
the RG transformation breaks down.
On the other hand, the boundary staggered field $h_B/J$
is still much smaller than 1 after the iterative RG transformations
because the boundary staggered magnetization is almost marginal.
Therefore, at the lowest order approximation, we may ignore the boundary
staggered field $h$.
Finally we reach the effective action of the boundary sine Gordon
model (5) together with the boundary condition (6), in the
main text.

\section{Boundary bound state}

Here, following Refs.~\cite{ghoshal_smatrix_suppl,ghoshal_bsg_suppl}, 
we briefly review the  energy spectrum of the boundary
sine Gordon model with the Neumann condition.
The energy spectrum in bulk is the same as those in the conventional
sine Gordon model.
A soliton (denoted as $S$) and an antisoliton (denoted as $\bar S$) 
exist and several breathers are formed as bound states of one soliton
and one antisoliton.
One can know that masses of breathers by analyzing simple poles of the
exact $S$ matrix.
As is well known, the $S$ matrix characterizes two-particle scatterings:
\begin{align*}
 Z_{a_1}(\theta_1)Z_{a_2}(\theta_2)
 &= S_{a_1a_2}^{b_1b_2}(\theta_1 - \theta_2)
 Z_{b_2}(\theta_2)Z_{b_1}(\theta_1) \\
 Z^\dagger_{a_1}(\theta_1) Z^\dagger_{a_2}(\theta_2)
 &= S_{a_1a_2}^{b_1b_2}(\theta_1 - \theta_2) Z^\dagger_{b_2}(\theta_2)
 Z^\dagger_{b_1}(\theta_1) \\
 Z_{a_1}(\theta_1)Z_{a_2}^\dagger(\theta_2)
 &=2\pi \delta_{a_1 a_2}\delta(\theta_1 - \theta_2) \\
 & \qquad+ S_{a_2b_1}^{b_2a_1}(\theta_1 - \theta_2)Z^\dagger_{b_2}(\theta_2)
 Z_{b_1}(\theta_1)
\end{align*}
Operators $Z_a(\theta)$ and $Z^\dagger_a(\theta)$ are called as
Faddeev-Zamolodchikov (FZ) operators.
The parameter $\theta$ is called as a rapidity, which parameterize the
energy $E_a$ and the momentum $P_a$ of a particle with an
index $a$ as
\[
 E_a(\theta) = M_a \cosh \theta, \quad P_a(\theta) = M_a \sinh \theta.
\]
Since every scattering in integrable field theories is factorizable to
several two-particle scatterings, the $S$ matrix determines whole energy
spectrum in bulk.
An $n$-th breather has a mass,
\begin{equation}
 M_n = 2M \sin \biggl( \frac{n\pi \xi}2\biggr),
  \label{eq.Mn}
\end{equation}
for $n = 1,2, \cdots, \lfloor \xi^{-1}\rfloor$.
Since breathers are formed by a soliton and an antisoliton with
the degenerate mass $M$, the mass of the breather $M_n$ is smaller than
$2M$.

\begin{figure}
 \centering
 \includegraphics[width=0.7\linewidth]{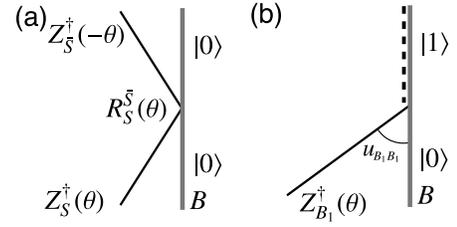}
 \caption{(a) A reflection process at the boundary.
 The boundary is represented as an infinitely heavy particle $B$ (thick
 gray line).
 (b) A formation of BBS using a breather $B_1$.
 The system is excited from the ground state 
 $|0\rangle = B|\mathrm{vac}\rangle$ to the
 one-BBS state $|1\rangle $.
 }
 \label{fig.diagram}
\end{figure}

Similar procedure is applicable to investigate the energy spectrum at
the boundary.
The scattering at the boundary is shown in Fig.~\ref{fig.diagram}~(a).
The spatial boundary is regarded as an infinitely heavy particle $B$.
\begin{equation}
 Z^\dagger_a(\theta)B = R_{\bar a}^b(\theta) Z_b^\dagger(-\theta)B
  \label{eq.scattering_boundary}
\end{equation}
The index $\bar a$ represents the anti-particle of the particle $a$.
The reflection factor $R_{\bar a}^b(\theta)$ has been studied in
detail~\cite{ghoshal_bsg_suppl}.
Simple poles of the reflection factor $R_{\bar a}^b(\theta)$ represents
energy levels localized at the boundary, called as a boundary bound
state (BBS).

For later convenience, we consider a rotated reflection factor
\begin{equation}
 K^{ab}(\theta) \equiv R_{\bar a}^b \biggl( \frac{i\pi}2 -
  \theta\biggr).
  \label{eq.Kab}
\end{equation}
According to Ref.~\cite{ghoshal_bsg_suppl}, the rotated reflection factor
$K^{ab}(\theta)$
has the following simple poles $\theta = iu_{ab}$ 
within the physical strip $0 <
\operatorname{Im}\theta < \pi/2$,
\begin{align}
 u_{ab} &= u_{SS} \equiv \pi \xi, \qquad (a,b=S,\bar S)
 \label{eq.uss} \\
 u_{ab} &= u_{B_1B_1} \equiv \frac{\pi}2 - \frac{\pi \xi}2, \qquad
 (a = b = B_1).
 \label{eq.ubb}
\end{align}
Here we assumed the experimentally realized 
situation $3 < \xi^{-1}<4$.
Both poles \eqref{eq.uss} and \eqref{eq.ubb} lead to the same BBS.
In fact, 
the excitation gap of the BBS, $M_{\mathrm{BBS}}$, is derived by the
 following two manners.
\begin{enumerate}
 \item One soliton $S$ or one antisoliton $\bar S$ forms a BBS:
       \begin{equation}
	M_{\mathrm{BBS}} = M \sin u_{SS} = M \sin (\pi \xi).
	 \label{eq.MBBS_ss}
       \end{equation}
 \item One first breather $B_1$ forms a BBS:
       \begin{equation}
	M_{\mathrm{BBS}} = M_1 \sin u_{B_1B_1}
	  = M \sin(\pi \xi).
	 \label{eq.MBBS_bb}
       \end{equation}
\end{enumerate}
In \eqref{eq.MBBS_bb}, we used the relation \eqref{eq.Mn}.
The formation of a BBS is sketched in Fig.~\ref{fig.diagram}~(b).
Note that the BBS is formed by a soliton with the mass $M$ or by a first
breather with the mass $M_1$.
The mass of the BBS should be smaller than $\min\{M, M_1\}$.
Since  $\min\{M, M_1\}$ is the lowest excitation gap, the BBS appears
below the bulk gap.

\section{Correlation functions}

We summarize correlation functions in the presence of the boundary.
In the Euclidean space, a two-point
correlation function $G_{S^+S^-}(x_1, x_2, \tau)$ depends on two spatial
variables $x_1$ and $x_2$ because the translational symmetry is broken
by the boundary (impurity).
For a systematic treatment, it is useful to make a Wick rotation
to regard the spatial variable $x$ as the ``time'', and
$\tau$ as the ``space'' variable.
In this framework,
\begin{equation}
 G_{S^+S^-}(x_1, x_2, \tau)
 = \frac{\langle \mathrm{vac}|T_x S^+(x_1,
  \tau)S^-(x_2, 0)|\mathcal B\rangle}{\langle \mathrm{vac}|\mathcal B
  \rangle}
  \label{eq.corr_real}
\end{equation}
The state $|\mathrm{vac}\rangle$ is
the ground state (with length given by
the inverse temperature of the original system,
and with the periodic boundary conditions).
The symbol $T_x$ denots the $x$ ordering, that is, the ordering which
orders operators with larger $x_i$ to right.
Note that the spin $S^a(x, \tau)$ is related to $S^a(0,0)$ via
\[
 S^a(x,\tau) = e^{-xE} e^{-i\tau P} S^a(0,0)e^{i\tau P}e^{xP}, 
\]
where $E$ and $P$ are the total energy and the total momentum
respectively. 
The state $|\mathcal B \rangle$ is called as a boundary
 state written in FZ operators and the rotated
 reflection factor~\cite{ghoshal_smatrix_suppl}.
\begin{align}
 |\mathcal B \rangle 
 &= e^K|\mbox{vac} \rangle \notag \\
 &= \exp \biggl[ \frac 12 \sum_{a,b} \int_{-\infty}^\infty
  \frac{d\theta}{2\pi} K^{ab}(\theta)Z^\dagger_a(-\theta)
  Z^\dagger_b(\theta)\biggr] |\mbox{vac} \rangle.
  \label{eq.boundarystate}
\end{align}

The Fourier transform of the $x$-ordered correlation function is given
 by
\begin{multline}
 G_{S^+S^-}(q, i \bar{\omega}) \\
  = \frac 1N \int_0^\infty d\tau \int_{-\infty}^0 dr
  dX \,
 e^{i(\bar{\omega} \tau - qr)} G_{S^+S^-}(x_1, x_2, \tau).
\end{multline}
where $\bar \omega$ is the Matsubara frequency.
Instead of spatial variables $x_1$ and $x_2$, we used
$r= -|x_1-x_2|$ and $X = (x_1+x_2)/2$.
The infinitely large $N$ represents the length of the semi-infinite spin
chain and the coefficient $1/N$ can be understood as a density of
non-magnetic impurities.
When we compute dynamical susceptibility, we perform the analytic
continuation  $i \bar{\omega} \to \omega + i \epsilon$,
where $\epsilon$ is positive infinitesimal.

\section{Bulk resonances}

Here we derive several resonant peaks in the ESR spectrum.
An $n$-particle state $|\theta_1, \theta_2, \cdots, \theta_n
\rangle_{a_1 a_2 \cdots a_n}$ and its conjugate
${}^{a_n\cdots a_2a_1}\langle \theta_n, \cdots, \theta_2, \theta_1|$
 are defined by
\begin{align*}
 |\theta_1, \theta_2, \cdots, \theta_n\rangle_{a_1a_2 \cdots a_n}
 &= Z^\dagger_{a_1}(\theta_1)Z^\dagger_{a_2}(\theta_2) \cdots
 Z^\dagger_{a_n}(\theta_n) |\mbox{vac} \rangle, \\
 {}^{a_n\cdots a_2a_1}\langle \theta_n, \cdots, \theta_2, \theta_1|
 &= \langle \mbox{vac}|Z_{a_n}(\theta_n) \cdots
 Z_{a_2}(\theta_2)Z_{a_1}(\theta_1).
\end{align*}
FZ operators $Z^\dagger_a(\theta)$ and $Z_a(\theta)$
respectively creates and annihilates a particle with a
rapidity $\theta$ with an index $a = S, \bar S, B_m$.
The energy $E=M_a \cosh \theta$ and the momentum $p = M_a \sinh \theta$ 
are parameterized by the single parameter $\theta$.
We put $v=1$ for simplicity.
The above multi-particle states are orthonormal.
\begin{multline}
 {}^{a_n \cdots a_1}\langle \theta_n, \cdots, \theta_1|
  \theta'_1, \cdots, \theta'_m \rangle_{a'_1 \cdots a'_m}\\
  = \delta_{nm} \delta_{a_1a'_1}\cdots \delta_{a_na'_n}
  (2\pi)^n \delta(\theta_1 - \theta'_1) \cdots \delta(\theta_n- \theta'_n)
\end{multline}
A set of $n$-particle states is complete:
\begin{widetext}
%
\begin{equation}
 1 = |\mbox{vac}\rangle \langle \mbox{vac}|
  + \sum_{n=1}^\infty \frac 1{n!} \sum_{a_1, \cdots, a_n}
  \int \frac{d\theta_1 \cdots d\theta_n}{(2\pi)^n}
   |\theta_1, \cdots, \theta_n \rangle_{a_1 \cdots a_n}{}^{a_n \cdots
  a_1}\langle \theta_n, \cdots, \theta_1|.
\end{equation}
This identity operator allows one to expand correlation functions by
intermediate states.
We can expand the boundary state in the power of $K$ as
\begin{equation}
 |\mathcal B \rangle =e^{K} |\mbox{vac}\rangle
 = |\mbox{vac} \rangle + K|\mbox{vac} \rangle + \cdots.
\label{eq.expansion_K}
\end{equation}

Let us consider the bulk part of the correlation
 $G_{S^+S^-}(q=\pi, i\omega)$, that is,
\begin{align}
 G^{(0)}_{S^+S^-}(q=\pi, i\bar{\omega})
 &= \frac 1N{C_s^\perp}^2 \int_0^\infty d\tau \int_{-\infty}^0 dr
 \int_{-\infty}^0 dX \,
  e^{i \bar{\omega}
 \tau}\langle \mbox{vac}|T_x e^{-i2\pi R \tilde
 \phi(x_1,\tau)} e^{i2\pi R \tilde \phi(x_2, 0)}|\mbox{vac} \rangle
 \notag \\
 &=\sum_{m=1}^{\lfloor \xi^{-1} \rfloor} \frac{v{C_s^\perp}^2}{M_m}
 |F_{B_m}^{e^{i2\pi R \tilde \phi}}|^2 \biggl( \frac 1{i \bar{\omega} - M_m}
 + \frac 1{i \bar{\omega} + M_m}\biggr) + \cdots
 \label{eq.G+-_pi_0_expanded} .
\end{align}
\end{widetext}

Thanks to the Lorentz invariance of the sine Gordon model,
the matrix element
$\langle \mbox{vac}|e^{-i2\pi R \tilde \phi(x_1,
\tau)}|\theta\rangle_a$
involving the intermediate state $|\theta\rangle_a$
can be computed easily.
\[
 \langle \mbox{vac}|e^{-i2\pi R \tilde \phi(x_1, \tau)}|\theta \rangle_{B_m}
  =  F_{B_m}^{e^{-i2\pi R \tilde \phi}} e^{i\tau M_m \sinh \theta +x_1
  M_m \cosh \theta}
\]
The factor $F_{B_m}^{e^{-i 2\pi R \tilde \phi}} \equiv \langle
\mathrm{vac}|e^{-i2\pi R \tilde \phi(0,0)}|\theta \rangle_{B_n}$
is a form factor of the operator $e^{-i2\pi R \tilde \phi}$.
Form factors generally depend on the rapidity $\theta$,
but the $F_{B_m}^{e^{-i2\pi R
\tilde \phi}}$ does not.
In integrable field theories, form factors can be exactly derived.
One can find several exact form factors of the sine Gordon model in
Ref.~\onlinecite{kuzmenko_DSF_suppl}.

After taking analytic continuation $i \bar{\omega} \to \omega + i \epsilon$, we obtain
\begin{multline}
 -\operatorname{Im}G^{(0)}_{S^+S^-}(q=\pi,  \omega+ i \epsilon) \\
=\sum_{n=1}^{\lfloor \xi^{-1} \rfloor} \frac{2\pi v
 {C_s^\perp}^2}{M_n} |F^{e^{i2\pi R \tilde \phi}}_{B_n}|^2 \delta
 (\omega - M_n) + \cdots.
 \label{eq.G+-_pi_0}
\end{multline}
Multi-particle resonant peaks such as $\delta (\omega - M_n - M_m)$
exist in the ignored terms.
\eqref{eq.G+-_pi_0} leads to peaks in \eqref{eq.mixing} as
\begin{equation}
 \biggl( \frac{D_z}J \biggr)^2
\sum_{n=1}^{\lfloor \xi^{-1} \rfloor} \frac{2\pi v
 {C_s^\perp}^2}{M_n} |F^{e^{i2\pi R \tilde \phi}}_{B_n}|^2 \delta
 (\omega - M_n) + \cdots.  
\end{equation}
As stated in the paper, the intensity of the staggered susceptibility
$\chi''_{+-}(q=\pi, \omega)$ is suppressed by the factor $(D_z/J)^2$.
Bulk parts of the resonant peaks which comes from 
$\chi''_{+-}(0, \omega)$ and $\chi_{zz}(\pi, \omega)$ are similarly
obtained, which are respectively
\begin{align}
  \frac{\pi v {C_u^\perp}^2}{M}|F_S^{e^{i\phi/R + i2\pi R \tilde
 \phi}}(\theta_0)|^2 \delta (\omega -E_S)&,
 \label{eq.G+-_0_0} \\
  \biggl( \frac{D_\perp}J\biggr)^2
 \frac{\pi v{C_s^z}^2}{M} |F_S^{e^{i\phi/R + i 2\pi R \tilde
 \phi}}(\theta_0)|^2 \delta(\omega - E_S)&.
 \label{eq.Gzz_pi_0}
\end{align}
$\theta_0 \equiv \ln \bigl[\bigl\{H+\sqrt{M^2+H^2}\bigr\}/M \bigr]$.
Both of them detect the soliton resonance at $\omega = E_S = \sqrt{M^2 +
H^2}$.
The latter \eqref{eq.Gzz_pi_0}
has a weaker intensity by the factor $(D_\perp/J)^2$ than the
former \eqref{eq.G+-_0_0}.
The $q=\pi$ component of $S^z_x$,
\[ 
 S^z_x \sim C_s^z(-1)^x \cos(\phi/R + Hx),
\]
is similar to the $q=0$ component of $S^\pm_x$,
\[
 S^\pm_x \sim C_u^\perp e^{\mp i 2\pi R \tilde \phi}
 \cos (\phi/R + Hx).
\]
Thus, all resonant peaks in $\chi''_{zz}(q=\pi, \omega)$ appear also in
$\chi''_{+-}(q=0, \omega)$, and the latter has stronger intensities in
\eqref{eq.mixing}.
We may forget $\chi''_{zz}(q=\pi, \omega)$ when we focus on 
resonance frequencies

We should emphasize that the breather resonances $\delta (\omega - M_n)$
can result only from the transverse staggered component
$\chi''_{+-}(q=\pi, \omega)$ in $\chi''_{\mathrm{phys}}(q=0, \omega)$.
This is an important consequence of the mixing \eqref{eq.mixing} caused
by the staggered DM interaction.
Without the mixing, we cannot observe any breather resonance.

\section{Boundary resonances}

\subsection{Transverse staggered component $G_{S^+S^-}(q=\pi, i\omega)$}

Boundary scatterings induce rich amounts of additional resonances.
For instance, we consider the first order contribution
in the expansion~\eqref{eq.expansion_K} to the $G_{S^+S^-}(q=\pi,i\omega)$,
\begin{widetext}
\begin{align}
 G^{(1)}_{S^+S^-}&(q=\pi, i\omega)  \notag \\
 &= \frac 1N {C_s^\perp}^2\int_0^\infty d\tau
 \int_{-\infty}^0 drdX \, e^{i\omega \tau}
 \frac 12 \sum_{a,b}
 \int_{-\infty}^\infty \frac{d\theta}{2\pi}
  K^{ab}(\theta) \langle \mbox{vac}|T_x e^{-i2\pi R
 \tilde \phi(x_1, \tau)}e^{i 2\pi R \tilde \phi(x_2, 0)} |-\theta,
 \theta \rangle_{ab}
 \label{eq.G+-_pi_1} \\
 &= \frac 1N {C_s^\perp}^2\int_0^\infty d\tau
 \int_{-\infty}^0 drdX \, e^{i\omega \tau}
 \frac 12 \sum_{a,b}
 \int_{-\infty}^\infty \frac{d\theta}{2\pi} K^{ab}(\theta)
 \sum_{n=1}^{\lfloor \xi^{-1} \rfloor}
 \int_{-\infty}^\infty \frac{d\theta_1}{2\pi} \notag \\
 & \quad \times \bigl[ \theta_H(x_2-x_1) \langle \mbox{vac}|e^{-i 2\pi R
 \tilde \phi(x_1, \tau)}|\theta_1 \rangle_{B_1}{}^{B_1}\langle \theta_1|
 e^{i2\pi R \tilde \phi(x_2, 0)}|-\theta, \theta \rangle_{ab} \notag \\
  & \qquad 
 + \theta_H(x_1 - x_2) \langle \mbox{vac}|e^{i 2\pi R \tilde
 \phi(x_2, 0)}|\theta_1 \rangle_{B_1}{}^{B_1}\langle \theta_1 |e^{-i
 2\pi R \tilde \phi(x_1, \tau)}|-\theta, \theta \rangle_{ab}
 \bigr] + \cdots
 \label{eq.G+-_pi_1_expanded}
\end{align}
$\theta_H(x)$ is the Heaviside step function.
As we reviewed above,
the factor $K^{ab}(\theta)$ has simple poles in the physical strip $0
\le \operatorname{Im} \theta < \pi/2$,
\begin{align}
 \mathop{\mathrm{Res}}_{\theta = i u_{SS}}
 K^{ab}(\theta)
 &= -if_S, \quad (a,b = S, \bar S),
 \label{eq.res_ss} \\
 \mathop{\mathrm{Res}}_{\theta = i u_{B_1B_1}}
 K^{ab}(\theta)
 &=- if_{B_1}, \quad (a = b = B_1),
 \label{eq.res_bb}
\end{align}
where $u_{SS}$ and $u_{B_1B_1}$ are \eqref{eq.uss} and \eqref{eq.ubb}.
The residues $-if_S$ and $-if_{B_1}$ are not derived at the moment of
 publication.
Here we assume $f_S, f_{B_1}>0$.
In the expansion of \eqref{eq.G+-_pi_1},
the most dominant intermediate state is one-breather state
$|\theta_1\rangle_{B_n}$.
The singularity of $K^{ab}(\theta)$ affects
\eqref{eq.G+-_pi_1} when and only when $n=1$ and $a=b=B_1$.
\begin{align*}
 G^{(1)}_{S^+S^-}(q=\pi, i\omega)
 &\sim \frac 1N {C_s^\perp}^2 \int_0^\infty d\tau \int_{-\infty}^0
 drdX \, e^{i\omega \tau} f_{B_1}|F^{e^{i2\pi R \tilde
 \phi}}_{B_1}|^2\\
 &\quad \times \bigl( e^{-\tau M_1 \sin u_{B_1B_1}}e^{2XM_1 \cos u_{B_1B_1}}
 +e^{\tau M_1 \sin u_{B_1B_1}}e^{2XM_1 \cos u_{B_1B_1}}
 \bigr)+ \cdots \\
 &=\frac{v{C_s^\perp}^2 f_{B_1} |F^{e^{i2\pi R \tilde
 \phi}}_{B_1}|^2}{2M_1 \cos u_{B_1B_1}} \biggl(\frac 1{i\omega - M_1 \sin
 u_{B_1B_1}} + \frac 1{i\omega + M_1 \sin u_{B_1B_1}}\biggr)+ \cdots
\end{align*}
The pole $ M_1 \sin u_{B_1B_1} = M \sin (\pi\xi) = M_{\mathrm{BBS}}$
gives a resonance of the boundary bound state (BBS) in
$\chi''_{\mathrm{phys}}(q=0, \omega)$:
\begin{equation}
 \biggl( \frac{D_\perp}J\biggr)^2 \frac{\pi v{C_s^\perp}^2 f_{B_1}
  |F^{e^{i2\pi R \tilde \phi}}_{B_1}|^2}{M_1 \sin (\pi \xi/2)}
  \delta (\omega - M_{\mathrm{BBS}})
  \label{eq.G+-_pi_1_bbs}
\end{equation}
The intensity of the peak \eqref{eq.G+-_pi_1_bbs} has two features.
(i) The intensity is suppressed by the factor $(D_\perp/J)^2$.
(ii) The intensity is independent of a concentration ($1/N$) of non-magnetic
impurities.
The first feature is common with the other resonant peaks which
originates in $\chi''_{+-}(q=\pi, \omega)$.
The second is also common with the other \textit{bulk} resonant peaks.
However, resonant peaks involved with the BBS generally depends on the
concentration.
For instance,
let us consider a two-particle intermediate state $|\theta_1, \theta_2
 \rangle_{B_1B_1}$ in $G^{(1)}_{S^+S^-}(q=\pi, i\omega)$ instead of
 the one-particle state $|\theta_1 \rangle_{B_1}$ in
 \eqref{eq.G+-_pi_1_expanded}.
The corresponding term in $G^{(1)}_{S^+S^-}(q=\pi, i\omega)$ is
\begin{multline*}
 \frac 1N {C_s^\perp}^2 \int_{0}^\infty d\tau \int_{-\infty}^0 drdX e^{i
  \omega \tau} \frac 12 \int_{-\infty}^\infty \frac{d\theta}{2\pi}
  K^{B_1B_1}(\theta) \frac 12 \int_{-\infty}^\infty \frac{d\theta_1
  d\theta_2}{(2\pi)^2} \\
 \quad \times
 \langle \mbox{vac}|e^{-i 2\pi R \tilde \phi(x_1, \tau)}|\theta_1,
 \theta_2 \rangle_{B_1B_1}{}^{B_1B_1}\langle \theta_2, \theta_1|e^{i2\pi
 R \tilde \phi(x_2, 0)}|-\theta, \theta \rangle_{B_1B_1}.
\end{multline*}
Here we assumed $x_1 < x_2$ for simplicity.
This process leads to a resonant peak in $\chi''_{\mathrm{phys}}(q=0, \omega)$
as follows.
\begin{equation}
 \frac 1N \biggl( \frac{D_z}J\biggr)^2 \frac{\pi v {C_s^\perp}^2
  f_{B_1}F_{B_1B_1}^{-i 2\pi R \tilde
  \phi}(-iu_{B_1B_1})F_{B_1B_1}^{i 2\pi R
  \tilde \phi}(-iu_{B_1B_1} + i\pi)}{2 M_1^2\sin(\pi \xi/2)}
  \delta (\omega - M_1 - M_{\mathrm{BBS}})
\end{equation}
The two-particle form factor
$ F_{B_1B_1}^{e^{i 2\pi R \tilde
 \phi}}(\theta_1 - \theta_2) \equiv \langle \mbox{vac}|e^{i 2\pi R
 \tilde \phi(0,0)}|\theta_1, \theta_2\rangle_{B_1B_1}
$
is positive when $\theta_1
 - \theta_2 = -iu_{B_1B_1}, -iu_{B_1B_1} + i\pi$~\cite{kuzmenko_DSF_suppl}.
Note that the intensity is proportional to the concentration of
 non-magnetic impurities because of the factor $1/N$.

The resonance frequency $\omega = M_1 + M_{\mathrm{BBS}}$ is a simple
 superposition of two resonances $\omega = M_1$ and $\omega =
 M_{\mathrm{BBS}}$.
If we consider $G^{(2)}_{S^+S^-}(q=\pi, i\omega)$, we will obtain
multi-BBS resonances such as $\omega = 2M_{\mathrm{BBS}}, M_n +
 2M_{\mathrm{BBS}}$.

\subsection{Transverse uniform component $G_{S^+S^-}(q=0, i\omega)$}

Below we point out that some boundary resonances cannot be understood by
any superposition of the existing resonance frequencies.
This new resonance appears in $\chi''_{+-}(q=0, \omega)$.
When we are concerned with $q\approx 0$ component of correlation
function $G_{S^+S^-}(q, i\omega)$, we may replace
\begin{equation}
 S^+_x \sim \frac{C_u^\perp}2 \bigl[ \mathcal O_S e^{-iHx} +
 \mathcal O_S^\dagger e^{iHx}\bigr].
 \label{eq.S2O}
\end{equation}
The fields
$ \mathcal O_S = e^{-i (2\pi R \tilde \phi + \phi/R)}$ and $\mathcal
O_S^\dagger =e^{i(2\pi R \tilde \phi + \phi/R)}$
can create $n$ solitons, $m$ antisolitons and $l$ breathers with $n-m =
1$ and $n-m = -1$ respectively.
Here $l$ can be arbitrary.
Thus, minimal intermediate states in $G_{S^+S^-}(q=0, i\omega)$ are a
one-soliton state $|\theta_1 \rangle_{S}$ and
a one-antisoliton state $|\theta_1 \rangle_{\bar S}$.

The uniform component $\chi''_{+-}(q=0, \omega)$
does not lead to the resonance at $\omega = M_{\mathrm{BBS}}$.
This is because a ``shift of the wavenumber'' is accompanied
 with $\chi''_{+-}(q=0, \omega)$.
The bulk soliton resonance 
occurs at $\omega =E_S= \sqrt{M^2 + H^2}$, not at
$\omega = M$.
$E_S$ is obtained by substituting $q=\pm H$ into the Lorentz invariant
dispersion relation
$E(q) = \sqrt{M^2 + q^2}$ of the soliton.
On the other hand, the BBS does not have $q$-dependent dispersion.
Thus, the resonance $\omega = M_{\mathrm{BBS}}$ cannot originate in
$\chi''_{+-}(q=0, \omega)$.

An interesting resonance appears in processes involved with
multi-particle intermediate states.
Let us consider a case in which a two-particle state $|\theta_1,
\theta_2 \rangle_{B_1S}$ appears as an intermediate state.
\begin{multline*}
 \frac 1N\frac{{C_u^\perp}^2}4 \int_0^\infty d\tau \int_{-\infty}^0 drdX \,
 e^{i\omega \tau} \frac 12 \sum_{a,b} \int_{-\infty}^\infty
 \frac{d\theta}{2\pi}K^{ab}(\theta)
 \int_{-\infty}^\infty\frac{d\theta_1d\theta_2}{(2\pi)^2} \\
 \quad \times \Bigl[ \theta_H(x_2-x_1) e^{-iHr} \langle \mbox{vac}|\mathcal
 O_S(x_1, \tau) |\theta_1, \theta_2 \rangle_{B_1S}{}^{SB_1}\langle
 \theta_2, \theta_1|
 \mathcal O_S^\dagger(x_2, 0) |-\theta, \theta
 \rangle_{ab} \\
  + \theta_H(x_1-x_2)e^{-iHr} \langle \mbox{vac}|\mathcal
 O_S(x_2, 0)|\theta_1, \theta_2 \rangle_{B_1S}{}^{SB_1}\langle \theta_2,
 \theta_1|\mathcal
 O_S^\dagger(x_1, \tau)|-\theta, \theta \rangle_{ab}
 \Bigr]
\end{multline*}
\end{widetext}
Both $B_1$ and $S$ in the intermediate state $|\theta_1, \theta_2
 \rangle_{B_1S}$
can form the BBS.
When the first breather $B_1$ forms a BBS, the resultant resonance
frequency is
\[
 \omega = M \cosh \theta_0 + M \sin u_{SS}.
\]
Since, $M\cosh \theta_0 =\sqrt{M^2+H^2}=E_S$ and $M \sin u_{SS} =
M_{\mathrm{BBS}}$, this resonance frequency is equal to
\begin{equation}
 \omega = E_S + M_{\mathrm{BBS}}.
  \label{eq.ES}
\end{equation}
This is a simple superposition of the bulk resonance $\omega = E_S$ 
and the boundary resonance $\omega = M_{\mathrm{BBS}}$
On the other hand, when the soliton $S$ forms a BBS, the resultant
 resonance frequency is somewhat strange, which is
\begin{equation}
 \omega = E_1+ M_{\mathrm{BBS}}.
  \label{eq.EB1}
\end{equation}
The first term is given by
\[
 E_1 = \sqrt{M_1^2 + H^2}.
\]
This resonance frequency \eqref{eq.EB1} cannot be represented by any
superposition of existing resonance frequencies.
Since a soliton $S$ is trapped at the boundary and transformed into a BBS,
a first breather $B_1$ propagates in the bulk 
as if it were a soliton.

\section{Intensities of ``unknown peaks''}

In our paper, we identified the ``unknown modes'' found in recent
experiments on KCuGaF$_6$~\cite{umegaki_KCuGaF6_suppl} and
Cu-PM~\cite{zvyagin_CuPM_prl_suppl,zvyagin_CuPM_suppl} with the boundary resonances.
Here we briefly comment with intensities of the ``unknown peaks''.
The compound KCuGaF$_6$ has larger staggered field ($c_s =0.18$ for $H
\parallel c$) than Cu-PM ($c_s = 0.083$).
Thus, in Cu-PM,
intensities of the staggered susceptibilities $\chi''_{+-}(q=\pi, \omega)$ and
$\chi''_{zz}(q=\pi, \omega)$ are strongly suppressed because
the strength of the staggered DM interaction is equal to $c_s \ll 1$ at most.
We may expect that, if boundary resonances exist, they are
likely to come from the uniform susceptibility $\chi''_{+-}(q=0,
\omega)$.
In fact, our assignments of the ``unknown modes'' in Cu-PM, which are
$\omega = E_{B_2}, E_{B_1} + M_1 - M_{\mathrm{BBS}}$, are consistent
with this speculation.
One can find that the breather resonances $\omega = M_n$ are weaker than
the ``unknown mode'' $U_1$ in Ref.~\onlinecite{zvyagin_CuPM_prl_suppl}.
We assigned $U_1$ to $\omega = E_{B_2}$.
While it looks counterintuitive that the boundary mode has the stronger
absorption intensities than those of bulk modes,
this relation can be understood by the weak mixing of the staggered
susceptibilities in \eqref{eq.mixing}.
The breather resonances come from the staggered part.

KCuGaF$_6$ has relatively stronger staggered field.
Ref.~\onlinecite{umegaki_KCuGaF6_suppl} shows that the breather resonances at
$\omega = M_n$ have approximately the same intensities as the soliton
resonances $\omega = E_S$.
This experimental observation is also consistent with \eqref{eq.mixing}
and the large $c_s$.
As for the ``unknown modes'', our assignments are consistent with the
estimation of the large $c_s$.
We assigned ``unknown modes'' to $\omega = M_{\mathrm{BBS}},
M_1 + M_{\mathrm{BBS}}, 2M_1-M_{\mathrm{BBS}}, M_2 - M_1
+M_{\mathrm{BBS}}$, all of which originate in the transverse part
$\chi''_{+-}(q=\pi, \omega)$.

\end{document}